\pdfoutput=1
\documentclass[sigconf,screen=true]{acmart}
\usepackage[T1]{fontenc}
\usepackage[utf8]{inputenc}

\usepackage{subcaption}
\usepackage{multirow}
\usepackage{rotating}
\usepackage{enumitem}
\setlist{nolistsep}

\acmSubmissionID{1365}

\setcopyright{rightsretained}

\begin{document}

\title{Integrated Speech and Gesture Synthesis}

\author{Siyang Wang}
\email{siyangw@kth.se}
\affiliation{
  \institution{Div.\ of Speech, Music and Hearing, KTH Royal Institute of Technology}
  \city{Stockholm}
  \country{Sweden}
}

\author{Simon Alexanderson}
\email{simonal@kth.se}
\affiliation{
  \institution{Div.\ of Speech, Music and Hearing, KTH Royal Institute of Technology}
  \city{Stockholm}
  \country{Sweden}
}

\author{Joakim Gustafson}
\email{jocke@speech.kth.se}
\affiliation{
  \institution{Div.\ of Speech, Music and Hearing, KTH Royal Institute of Technology}
  \city{Stockholm}
  \country{Sweden}
}

\author{Jonas Beskow}
\email{beskow@kth.se}
\affiliation{
  \institution{Div.\ of Speech, Music and Hearing, KTH Royal Institute of Technology}
  \city{Stockholm}
  \country{Sweden}
}

\author{Gustav Eje Henter}
\email{ghe@kth.se}
\affiliation{
  \institution{Div.\ of Speech, Music and Hearing, KTH Royal Institute of Technology}
  \city{Stockholm}
  \country{Sweden}
}

\author{\'{E}va Sz\'{e}kely}
\email{szekely@kth.se}
\affiliation{
  \institution{Div.\ of Speech, Music and Hearing, KTH Royal Institute of Technology}
  \city{Stockholm}
  \country{Sweden}
}


\begin{abstract}
Text-to-speech and co-speech gesture synthesis have until now been treated as separate areas by two different research communities, and applications merely stack the two technologies using a simple system-level pipeline.
This can lead to modeling inefficiencies and may introduce inconsistencies that limit the achievable naturalness.
We propose to instead synthesize the two modalities in a single model, a new problem we call integrated speech and gesture synthesis (ISG).
We also propose a set of models modified from state-of-the-art neural speech-synthesis engines to achieve this goal.
We evaluate the models in three carefully-designed user studies, two of which evaluate the synthesized speech and gesture in isolation, plus a combined study that evaluates the models like they will be used in real-world applications -- speech and gesture presented together. The results show that participants rate one of the proposed integrated synthesis models as being as good as the state-of-the-art pipeline system we compare against, in all three tests.
The model is able to achieve this with faster synthesis time and greatly reduced parameter count compared to the pipeline system, illustrating some of the potential benefits of treating speech and gesture synthesis together as a single, unified problem.

\end{abstract}


\copyrightyear{2021} 
\acmYear{2021} 
\acmConference[ICMI '21]{Proceedings of the 2021 International Conference on Multimodal Interaction}{October 18--22, 2021}{Montréal, QC, Canada}
\acmBooktitle{Proceedings of the 2021 International Conference on Multimodal Interaction (ICMI '21), October 18--22, 2021, Montréal, QC, Canada}
\acmDOI{10.1145/3462244.3479914}
\acmISBN{978-1-4503-8481-0/21/10}


\settopmatter{printfolios=true} 
\maketitle

\section{Introduction}
Humans make use of both verbal and nonverbal communication to achieve efficient, expressive, and robust face-to-face interaction.
Both are fundamentally intertwined, born out of a common representation of the message to be communicated, colored by the situation at hand. 
Interactions with social robots and embodied conversational agents (ECAs) would benefit from complementing their speech with nonverbal communication like co-gestures \cite{bergmann2013virtual,luo2013examination,wu2014effects}. Existing approaches to simultaneous speech and gesture generation have so far simply combined disjunct speech-synthesis systems with gesture-generation components that are trained separately.
In this paper, we investigate a fully integrated approach, which we call \emph{integrated speech and gesture synthesis} (ISG).

\begin{figure}[!t]
  \centering
  \subcaptionbox{Pipeline\label{fig:pipeline}}{%
      \includegraphics[height=.48\linewidth]{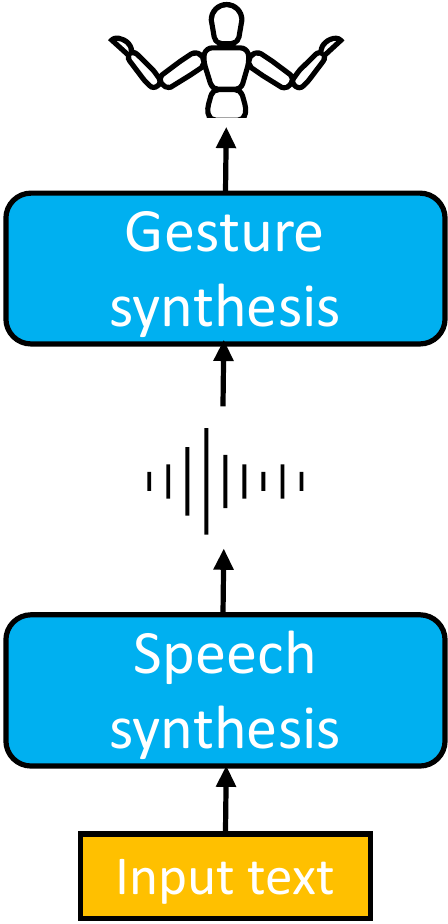}}%
  \hspace{.1\linewidth}
  \subcaptionbox{Integrated\label{fig:co-speech-gesture}}{%
      \includegraphics[height=.48\linewidth]{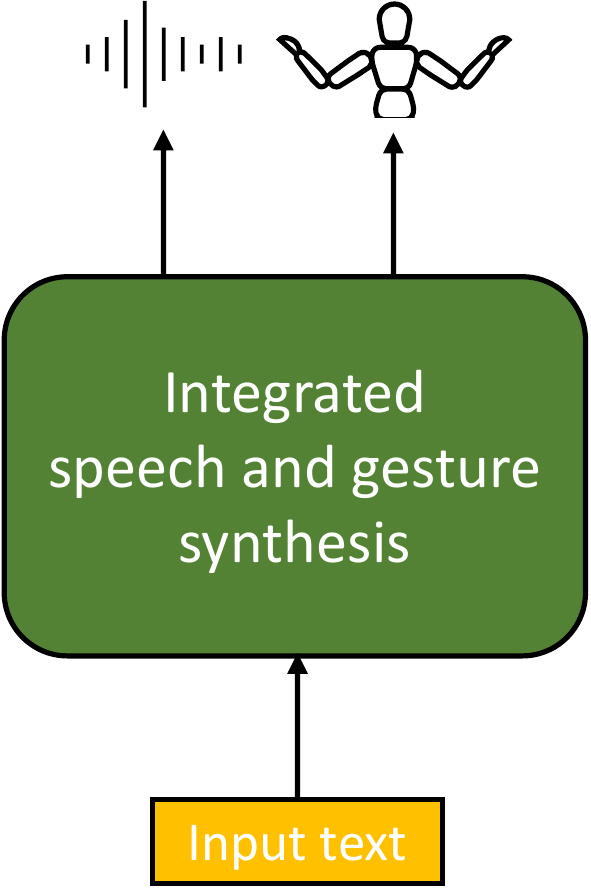}}%
  \caption{Two paradigms for speech and gesture synthesis.}
  \label{fig:splash}
\end{figure}

A common approach to obtaining a speaking and gesturing agent is to stack speech- and gesture generators in a pipeline as shown in Figure \ref{fig:pipeline}.
The pipeline generates speech audio from input text using the speech-synthesis component, and then passes that audio to the gesture-generation component to generate matching gestures \cite{alexanderson2020generating}.
Gesture-generation systems can also be driven by text input, which is common for rule-based systems (see \ref{ssec:motionsynthesis}), but for gesticulation and speech to remain synchronized for these agents, it is still necessary for these systems to leverage information from the speech audio, such as word-level timing information (e.g., \cite{cassell2001beat,yoon2019robots}).
This again implies a pipelined approach where speech is synthesized first.

Both speech synthesis from input text and gesture synthesis from input speech audio are well-studied problems on their own.
The state-of-the-art in both areas are data-driven models with deep-learning backbones trained on large datasets.
A pipelined text-to-speech then speech-to-gesture approach allows the two components to be trained separately, and potentially on different datasets.
The approach also generally benefits from improvement in either component.
However, the pipeline approach also has a number of notable drawbacks:
First, the output from the speech-synthesis module is usually not of the same quality -- most often not even the same voice -- as the ground-truth speech audio that the gesture-generation module is trained on.
This may degrade the quality of the generated gestures as a result.
Second, training two systems separately is inefficient.
A typical neural network-based gesture-generation model that takes speech audio as input first needs to extract features from the speech audio, such as the duration, pitch, and intensity of phonemes, in order to generate gestures that match the speech \cite{kucherenko2021large,alexanderson2020style}.
However, such features are already explicitly or implicitly modeled within speech-synthesis models \cite{shen2018natural}, and forcing the gesture-generation component to extract features from speech audio that already existed inside the speech-synthesis component is suboptimal.

The goal of this paper is to unify speech and gesture generation, in order to bring together the separate research communities of TTS and co-speech gesture synthesis.
Our main contributions are:
\begin{itemize}
\item We pose and explore the novel problem of building multimodal systems that jointly synthesize speech and gestures in a single, integrated deep-learning architecture.
\item We propose two sets of gesture-generation systems based on two representative neural speech-synthesis architectures, namely Tacotron 2 \cite{shen2018natural} (deterministic, autoregressive) and Glow-TTS \cite{kim2020glow} (probabilistic, parallel).
\item We identify previously-unknown design challenges and trade-offs faced when creating and evaluating integrated systems.
\item We evaluate the proposed speech-and-gesture synthesis models in depth against a state-of-the-art pipeline system \cite{alexanderson2020generating}.
\end{itemize}
Our evaluation considers both synthesized gesture and speech in isolation, along with a third test evaluating time-aligned gesture and speech together. The combined results show that one proposed ISG model achieves same performance as the state-of-the-art pipeline system with 3.5 times fewer parameters and faster synthesis time.
The speech and gesture from all evaluated models is included in the supplement.
For code and video please see \href{https://swatsw.github.io/isg_icmi21/}{our project page}.\footnote{\href{https://swatsw.github.io/isg_icmi21/}{https://swatsw.github.io/isg\_icmi21}; videos are also in supplemental materials.}

\section{Background}
Historically, the synthesis of speech and gesture have been treated as different problems, approached with different goals and different types of data by often disjunct research communities.
Lately, however, both fields have moved towards ever more domain-agnostic machine-learning technologies.
This section discusses where the two fields are today and highlights recent convergent trends.

\subsection{Speech synthesis}
\label{ssec:tts}
State-of-the-art speech synthesis is largely deep learning-based, both for waveform modeling (neural vocoders) following WaveNet \citep{oord2016wavenet}, and sequence-to-sequence approaches for acoustic modeling (spectrogram generation), first seen in \cite{wang2016first} and later popularized by Tacotron \citep{wang2017tacotron}.
These two trends were brought together in Tacotron 2 \cite{shen2018natural}, which established a new state of the art by separately training a seq-to-seq acoustic model to generate mel spectrograms from text and a neural vocoder to synthesize waveforms from the mels.
This is the dominant approach today.
The rise of normalizing flows like Glow \citep{kingma2018glow} for neural vocoders \citep{prenger2019waveglow,kim2019flowavenet,ping2020waveflow} and acoustic models \citep{valle2020flowtron,miao2020flow,kim2020glow}
has created strong probabilistic TTS models.
These can avoid the averaging artifacts seen with deterministic approaches,
such as flat intonation and reduced speaker similarity due to over-smoothing.
In this paper, we take two leading TTS architectures, one deterministic (Tacotron 2 \cite{shen2018natural}) and one probabilistic (Glow-TTS \cite{kim2020glow}), and use them as a starting point for new architectures that generate both speech acoustics and body poses at the same time.

Neural TTS approaches have also seen success in synthesizing convincingly spontaneous-sounding speech \citep{szekely2019spontaneous} with filled pauses \citep{szekely2019how} and breathing \citep{szekely2020breathing}.
This ability to synthesize convincing spontaneous speech from text is a key enabler of integrated speech and gesture synthesis,
since spontaneous speech generally is accompanied by gestures, while speech read aloud from text (used in the vast majority of TTS training corpora) is not.

\subsection{Gesture generation}
\label{ssec:motionsynthesis}
Co-speech gesture generation has traditionally been dominated by rule-based systems (e.g., \cite{cassell1994animated,cassell2001beat,kopp2004synthesizing,ng2010synchronized,marsella2013virtual}; see \cite{wagner2014gesture} for a review). The use of machine learning to generate gestures is relatively more recent, and there are also hybrid systems that combine learned and procedural approaches, e.g., by learning when to produce a gesture from a fixed set of pre-animated gesture clips \cite{kipp2005gesture,neff2008gesture,ishi2018speech,chiu2015predicting}.

Among systems that leverage machine-learning for speech-driven gesture generation, one can make a distinction based on the modality used to represent the input speech: either audio, text, or both \cite{kucherenko2021multimodal}.
Audio-based gesture-generation systems include \cite{hasegawa2018evaluation,kucherenko2021moving,ginosar2019learning,ferstl2020adversarial,henter2019moglow,lu2021double}.
Systems of this kind usually generate mainly beat gestures (gestures that align with the rhythm of the speech) and are a natural fit for use in pipeline approaches to speech-and-gesture generation.
Text-based systems, on the other hand, are seen as better suited for generating representational and communicative gestures.
However, even text-based gesture-generation systems that lack audio as an explicit input, e.g., \cite{ishi2018speech,yoon2019robots}, typically still require word-level timing information to synchronize gestures and speech.
This information is not available from text, but only from audio, or from the process that creates it.
Finally, in recognition of the fact that both acoustic information (from the speech audio) and semantic information (from a text transcription) are complementary for the task of generating communicative and natural-looking speech-driven gestures, the field is experiencing a rapid shift towards methods that use both audio and text inputs simultaneously \cite{kucherenko2020gesticulator,yoon2020speech,ahuja2020no,korzun2020finemotion}.
This trend suggests that our proposal for integrated speech-and-gesture synthesis, where the generated gestures may be informed by both text and acoustic properties, is timely and worth pursuing.

\subsection{Towards integrated multimodal synthesis}
\label{ssec:joint}
Many embodied agents leverage speech and gesture synthesis components in a pipeline approach. However, the vast majority of these agents use an incoherent setup where the speech synthesis is trained on a different dataset than the gesture generation, and style and speaker identity may differ between components. In fact, the only system we are aware of where both components explicitly were trained on the same dataset is the one in \cite{alexanderson2020generating}, and we consequently use their approach as the baseline pipeline approach for our experiments in Section \ref{sec:experiments}.

In terms of multimodal synthesis beyond speech-and-gesture, the most similar work to ours that we are aware of is DurIAN \cite{yu2019durian}.
Our work differs in three aspects:
First, and most important, DurIAN aims to co-generate facial expression and speech, instead of gesture and speech. Gesture generation and facial-expression synthesis are different problems, evident in that they have attracted separate research communities.
Synthesizing gesture and speech in a single model is a novel problem that, to our best knowledge, we are the first to study. 
Second, we use different speech-synthesis frameworks from DurIAN.
Third, we use a dataset of spontaneous speech and 3D motion capture from a single speaker, which is very different from the dataset used in DurIAN.

\begin{figure}[!t]
  \centering
  \includegraphics[width=\linewidth]{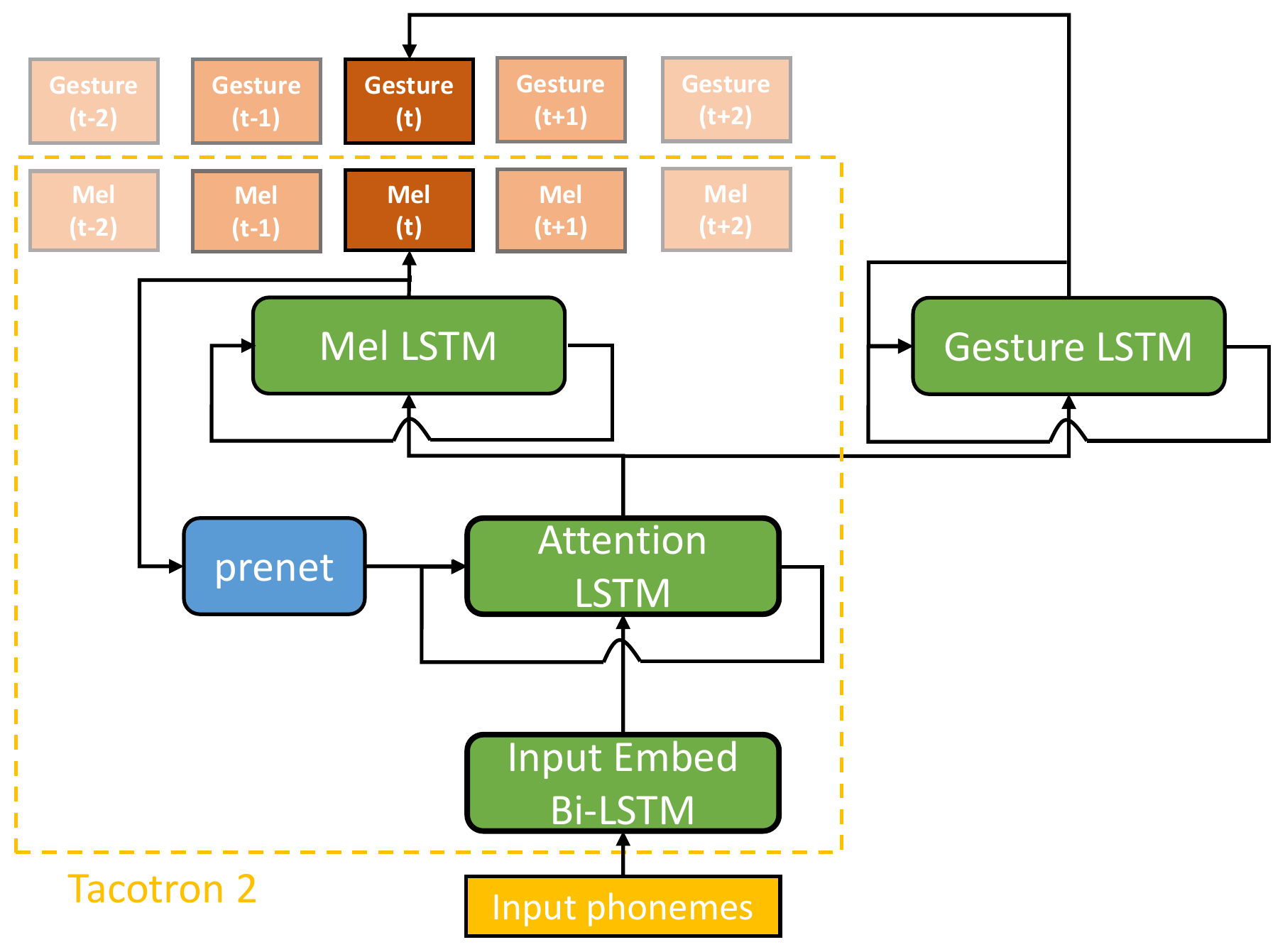}
  \caption{Proposed Tacotron2-based integrated speech and gesture generation model (Tacotron2-ISG). The Tacotron 2 residual conv.\ postnet that post-processes the mel spectrogram output is also in Tacotron2-ISG but not shown here.}
  \Description{.}
  \label{fig:modified_tacotron2}
\end{figure}
\section{Integrated speech and gesture synthesis}

\subsection{Problem definition}
We define a paired speech-gesture dataset as one that records a human actor gesturing and speaking at the same time and the two modes are time-aligned in the dataset.
The speech audio is usually a waveform and the gesture is usually in fixed-frame-rate joint angles or joint positions.
Given such a dataset, the problem is to create a machine-learning model that takes text 
as input and generates both speech audio and gesture. 
This is an ill-defined problem similar to speech synthesis, due to the fact that the input (text) has much lower information rate than the output (speech and gesture).
This means that we have to carefully define the goal for a successful model.
While objective measures like mean square error on a held-out set are helpful, what we really care about is how the model is perceived by human users in an interaction.
We therefore evaluate our models in subjective tests, detailed in Sections \ref{sec:experiments} through \ref{sec:results}.

\subsection{Proposed models}
We develop our ISG models starting from two state-of-the-art TTS models, Tacotron 2 \cite{shen2018natural} and Glow-TTS \cite{kim2020glow}. These represent two main approaches to TTS, with Tacotron 2 being auto-regressive and non-probabilistic and Glow-TTS being parallel and probabilistic. The adaptation process detailed below is largely guided by experimentation along with a hypothesis about how to best utilize representations inside TTS frameworks for gesture generation. 

\subsubsection{Tacotron 2-based models}
\label{section:tacotron2-based-models}
We adapt Tacotron 2 for ISG by adding a gesture generation sub-network to the architecture and employing different training strategies such as transfer learning, parameter freezing, and an adversarial loss. We focus on two goals, (a) utilizing the intermediate representations in Tacotron 2 to generate gesture and (b) changing as little of the original architecture as possible. 
We hypothesize that the attention-LSTM layer representation in Tacotron 2 is the most useful for gesture generation because it should correlate with high-level speech-planning.
Hence we use this representation as input to the gesture generation sub-network.

We are careful about changing the original speech-synthesis model, because early experiments we conducted found that even minor changes to Tacotron 2 affect the TTS quality.
This is especially true for the attention layer:
Learning to attend is usually a training bottleneck and attention failures such as skipping and babbling are a weak point of the Tacotron architecture \cite{watts2019where,battenberg2020location}.
This issue is especially pronounced when training on small databases \cite{xu2020lrspeech}, and currently-available corpora of spontaneous speech and 3D gesture are often smaller than standard corpora used to train TTS.
The monotonic attention mechanism of Glow-TTS \cite{kim2020glow} is one solution to this problem (among others, e.g., \cite{yu2019durian,battenberg2020location,xu2020lrspeech,shen2020non}).
We tried to add the generated gesture to the input of the prenet which outputs to the attention-LSTM layer, but we found that this small change causes speech synthesis to get much worse.
For these reasons, we decide to keep all original Tacotron 2 modules intact (including original hyperparameters \cite{shen2018natural}), and combine Tacotron 2 with the gesture generation sub-network by only using the attention-LSTM layer output as input to the completely separate gesture-generation module as described above and shown in Figure \ref{fig:modified_tacotron2}.

We take a fine-tuning approach to train the model.
This is mainly because Tacotron 2 and especially its attention LSTM layer is known to be difficult and slow to train from ground up.
In fact, it is not uncommon in the speech synthesis community to use transfer learning to train a Tacotron 2 pre-trained on large scale read speech dataset to a smaller speech dataset \cite{szekely2019spontaneous}.
We take this approach even further, by first taking a Tacotron 2 model pre-trained on read speech and training it on our speech data only without gestures, and then adding the gesture sub-network for ISG training.

During the ISG training stage, we also experiment with freezing the weights in the speech sub-network, which prevents the possibility that ISG training focuses on improving gesture at the expense of already achieved speech quality.
However, this approach takes away supervision from the speech loss which makes it more challenging for the gesture sub-network to generate speech-aligned gesture. Furthermore, the prenet of Tacotron 2 shown in Figure \ref{fig:modified_tacotron2}, has a random dropout of 0.5 at both train and test time. This means that the attention LSTM layer output varies for the same input text even with weights frozen, making it difficult for the gesture sub-network to regress to the same ground-truth gesture while its input, taken from the output of the attention LSTM layer of Tacotron 2, is constantly changing. 
This is in contrast to conventional speech to gesture setups, where the input is a finite amount of constant ground-truth speech from the dataset and the model is regressing to the corresponding gesture.
On the other hand, this noisy input is very similar to GANs, in which the generator receives random noise input and generates convincing samples by optimizing itself against an adversarial discriminator.
Thus, in the model where the speech sub-network is frozen during ISG training, we add a discriminator that takes in both generated speech and gesture as input to encourage the generated gesture to align with generated speech.
Inputting both speech and gesture to the discriminator is a practice that has been shown to be effective in encouraging speech-gesture consistency by previous studies \cite{habibie2021learning,ferstl2019multi}. However, an important distinction of ours is that the speech in our model is also generated instead of using ground-truth speech as in those studies.

\subsubsection{Glow-based model}
We also developed a model building on the recent flow-based TTS system called Glow-TTS \cite{kim2020glow}.
Unlike the autoregressive and LSTM-based architecture of Tacotron 2, Glow-TTS acts in parallel using convolution operators, making it faster to synthesize from on GPUs and avoids stability issues (feedback loops) that are a risk in autoregressive models.
We were also interested in the probabilistic aspects of flow-based architectures, since gesture realizations show great variability for a given utterance, and a deterministic summary such as the average gesture cannot capture that diversity.

Glow-TTS has three main components: A transformer-based text encoder $f_\mathrm{{enc}}$, a flow-based convolutional decoder $f_{\mathrm{dec}}$, and a text-to-speech aligner $A$ that time-aligns the input and output, i.e., that associates each mel-spectral frame of speech acoustics with an embedding produced by the encoder.
During training, the normalizing flow in the decoder invertibly transforms the mel-spectrogram representation $x$ of a given utterance to a latent representation $z=g(x|c)$, conditioned on the text $c$.
(The invertibility of the transformation $g$ means that we can use the transformation of variables formula to compute the data likelihood. \cite{kingma2018glow})
By letting the latent distribution be elementwise Gaussian and parameterizing it with the alignment $A$ and means and standard deviations $\mu, \sigma = f_{\mathrm{enc}}(c)$ returned by the decoder, one obtains a probabilistic model $P_X(x|c)$ of the original speech given the text. The model can then be optimized by gradient-based methods using a combination of maximum-likelihood estimation and Viterbi decoding.
During synthesis, $f_{\mathrm{enc}}(c)$ and $A$ are used to predict speech-sound durations and then sample a sequence of conditional random latents from $P_Z(z|c)$ that conforms to these durations.
The decoder then transforms these sampled latents into mel-spectral frames, thus obtaining a
sample from the distribution $P_X(x|c)$. The resulting mel-spectrogram $x$ is passed to a vocoder (HiFi-GAN \cite{kong2020hifi}) to generate a speech waveform.

In this work, we let the decoder model the multimodal distribution $P_X(x_a, x_m|c)$, where the subscripts $a$ and $m$ denote \emph{audio} and \emph{motion}, by simply extracting audio and motion features at the same time instances (same frame rate) and concatenating them into a unified vector for each frame.
Glow-TTS being a normalizing flow model, its layer-wise representation is entangled thus making it difficult to use any one layer as input for generating gesture as was done for Tacotron2-ISG. 

\section{Data}
\subsection{Training corpus}
For the experiments we used the recordings from the Trinity Speech-Gesture Dataset \cite{IVA:2018} as processed by \cite{kucherenko2021large}.
The dataset comprises 25 impromptu monologues by a male speaker of Irish English, on average 10.6 minutes long, from a multi-camera motion-capture studio.
The actor speaks in a colloquial style, spontaneously and without interruption on topics such as hobbies and interests.
During the monologues, he addresses a person seated behind the cameras who is giving visual, but no verbal feedback. 

To create a corpus suitable for speech synthesis, the audio recordings were automatically segmented into breath groups with an automatic breath detection method \citep{szekely2019casting}. 
Words were transcribed using Google Cloud Speech-to-Text API and subsequently manually corrected to ensure that all words were accurate.
All speech events were transcribed using ARPABET phones, no new characters were introduced outside of the standard.
In order to maximize the utterance length in the corpora and to enable insertion of inhalation breaths in the TTS, we used a data augmentation method called \emph{breathgroup bigrams}.
This method essentially consists of segmenting a speech corpus into stretches of speech delineated by breath events, and then combining these breath groups in an overlapping fashion to form utterances no longer than 11 seconds \cite{szekely2020breathing}.
This method also makes it possible to leverage the continuous nature of the recordings and learn contextual information beyond respiratory cycles.
The minor misalignment between motion and speech in the original dataset was manually corrected. The motion is represented by exponential map of joint rotations \citep{grassia1998practical}.
We only used the upper body data and removed the fingers (but not hand orientation) due to low capture accuracy there.
For visualization, we instead used fixed, lightly-cupped hands on the avatar, similar to \cite{alexanderson2020generating}.

\subsection{Text inputs for the evaluation}
Selecting the input text for evaluating spontaneous speech synthesis is not straightforward, because the training data does not conform to the conventions of written language and lacks a clear sentence structure \cite{szekely2020augmented}.
In this work, the evaluation uses text prompts that were semi-automatically generated by the medium-size pre-trained GPT-2 model \cite{radford2019language} (355 million parameters) fine-tuned on the TTS corpus.
Since there are no sentence boundaries in the spontaneous speech corpus, commas indicating pauses and periods were used as breath tokens and end-of-utterance markers. The last breath token was replaced by an end-of-utterance marker.
When generating the prompts, common first-person phrases were used as prefixes to seed the GPT-2 samples, and these prefixes were included in the generated prompt.
The generated texts underwent a manual selection process to identify 17 semantically coherent utterances 
mostly between 25 and 50 tokens long.
We find that longer sentences are more suitable for evaluating ISG because they differentiate models more than shorter sentences do, since models have to generate more gesture strokes while being consistent. 
Using GPT-2-generated input sentences means that we do not have access to corresponding ground-truth speech and gesture.
This makes it difficult to assess the gap between the synthesis and the ground truth, but addressing this question is not the main purpose of this study. In addition, using generated sentences allows us to evaluate model performance ``in the wild'' to some degree.

\section{Evaluation}
\label{sec:experiments}
\subsection{Model and training configurations}
\label{section:model_training_config}
\subsubsection{Baseline: Pipeline with Tacotron 2 and StyleGestures}
We compare the models against a state-of-the-art pipeline system \cite{alexanderson2020generating} which uses Tacotron 2 for speech synthesis and StyleGestures \cite{alexanderson2020style} for gesture generation.
The only changes we made are using WaveGlow \citep{prenger2019waveglow} for vocoding, as opposed to the Griffin-Lim algorithm used in \cite{alexanderson2020generating}, and only training on upper-body data.
We used the publicly available, official implementation of StyleGestures\footnote{\href{https://github.com/simonalexanderson/StyleGestures}{https://github.com/simonalexanderson/StyleGestures}}, using the recommended hyperparameters with 16 flow-steps and 512 channels in the LSTM layers.
The model was trained for 80k iterations.
We initialized the autoregressive context with a static mean pose and padded the test audio with 1 s of silence at the end to account for the future speech data input the gesture-generation model requires.   

\subsubsection{Tacotron2-ISG}
We use an open-source Tacotron 2 repository \footnote{\href{https://github.com/NVIDIA/tacotron2}{https://github.com/NVIDIA/tacotron2}} and the pre-trained model that it comes with. We also use the pre-trained WaveGlow model in that repository for vocoding. Unless stated otherwise, the hyper-parameters used in both speech-only training and ISG training are the same as the repository defaults.
We then add the gesture sub-network consisting of 4 LSTM layers, each with 512 nodes, the input of which is the attention-LSTM layer output of Tacotron 2 as shown in Figure \ref{fig:modified_tacotron2} together with the prior frame gesture output for autoregressive learning.
The resulting model, which we call \emph{Tacotron2-ISG}, is then trained for integrated speech-gesture generation in two ways, (a) with both speech and gesture sub-networks trained simultaneously, which we call Co-training Tacotron2-ISG or \emph{CT-Tacotron2-ISG}, (b) the speech sub-network weights frozen after adding the gesture sub-network, and training only gesture sub-network, which we call Separate-Training Tacotron2-ISG or \emph{ST-Tacotron2-ISG}.
In all cases the training loss is mean squared error (MSE), but as mentioned in Section \ref{section:tacotron2-based-models}, the ST-Tacotron2-ISG model (in which the speech sub-network is frozen) also has an added GAN loss on the combined speech-gesture output, so that the speech sub-network can provide additional supervision despite being frozen.
The GAN loss is in its original form \cite{goodfellow2014generative} and is weighted by 0.05, as we found that too strong a GAN loss results in bad gestures.
We use a discriminator consisting of 2 LSTM layers, each with 1024 nodes. 
We apply scheduled sampling \cite{bengio2015scheduled} when training the gesture sub-network, where the teacher forcing probability is 1 for the first 5 epochs, then linearly drops to 0.2 over 40 epochs,
remaining at that value for the rest of the training.
We do not use scheduled sampling to train the speech sub-network because Tacotron 2 already works well with full teacher forcing.
The gesture-generation sub-network is also autoregressive as it takes the prior frame (pose) as input, in addition to the output of the attention-LSTM in the speech sub-network.
We find that the gesture sub-network works best running at 20 fps.
However, the attention-LSTM layer of the speech sub-network (Tacotron 2) runs at roughly 80 fps.
Thus, the gesture sub-network takes output from the attention-LSTM layer every 4 frames to match its own 20 fps rate.
In order to match the gesture smoothness of the other models, we also apply 1-D Gaussian filters to the generated gesture in sliding windows with stride 1 and window size 3.

\begin{table}[!t]
   \caption{Model parameter counts and average synthesis time with 95\% confidence intervals.}
  \label{tab:parameter_count}
\begin{tabular}{@{}lcc@{}}
\toprule 
 System & Param.\ count & Synth.\ time\tabularnewline
\midrule
 Pipeline \cite{alexanderson2020generating}, comprising
2 sub-systems: & 137.53M & 5.08$\pm$0.49 s\tabularnewline
$\quad$ TTS: Tacotron 2 \cite{shen2018natural}  & \hphantom{0}28.19M & 1.56$\pm$0.15 s\tabularnewline
$\quad$ gesture: StyleGestures \cite{alexanderson2020style} & 109.34M & 3.52$\pm$0.34 s\tabularnewline
\midrule
 Tacotron2-ISG (ours) & \hphantom{0}38.83M & 1.49$\pm$0.13 s\tabularnewline
 GlowTTS-ISG (ours) & \hphantom{0}28.95M & 1.64$\pm$0.12 s\tabularnewline
\bottomrule
\end{tabular}
\end{table}

\subsubsection{GlowTTS-ISG}
For the modified Glow-TTS model we used a temporal resolution of 60 fps for both the audio and motion features, and re-trained the HiFi-GAN vocoder \cite{kong2020hifi} to generate audio at this frame rate.
For model training, we used the same hyper-parameters as \cite{kim2020glow}, except for the number of blocks in the one-by-one (depth-wise) convolution in each flow layer, where we used 25 blocks of 10 features instead of the original 40 blocks of 4 features.
This results in slightly longer training and synthesis time but ensures a more expressive model through better channel mixing.
We followed the recommended procedure of adding blank spaces between each text-token and trained the model for 275k iterations.
We then generated evaluation samples with a temperature of 0.7 and length scale of 0.9, providing a good mix between output variation and naturalness.
We call this model \emph{GlowTTS-ISG}.

\subsection{Model size and synthesis time}
Table \ref{tab:parameter_count} reports the number of parameters of the compared models. 
Both ISG models have comparable parameter count, and have at least 3.5 times fewer parameters than the pipeline system. If either ISG model obtains just same level of perceptual evaluation results as the pipeline system, we can say that the ISG model is more parameter-efficient.
Table \ref{tab:parameter_count} also shows that the pipeline model has the longest synthesis time, both as a result of its sequential nature (i.e., the gesture sub-system cannot start running before the TTS completes) and due to the computation-heavy gesture sub-system StyleGestures.\footnote{The results we report here are based on a modification of the official implementation from \href{https://github.com/simonalexanderson/StyleGestures}{https://github.com/simonalexanderson/StyleGestures}, to
cache the inverse matrix computations in the flow. This sped up generation time by approximately a factor five.}
Tacotron2-ISG is faster since it eliminates the latency between TTS and gesture synthesis and has a more efficient gesture module. GlowTTS-ISG has comparable synthesis time to Tacotron2-ISG on the tested inputs, but is expected be faster on longer inputs due to its non-autoregressive design.
The time used by the vocoder (WaveGlow or HiFi-GAN) is not included in the measurements, since either system can be used with either vocoder. To ensure that the difference in synthesis times reflects different complexity of the models, and not length differences in the synthesized utterances, we calculated the mean utterance duration for the evaluated systems, finding them to be roughly the same: 8.33 s (Pipeline), 8.26 s (Tacotron2-ISG), and 9.76 s (GlowTTS-ISG).

\begin{figure}[!t]
  \centering
  \includegraphics[width=\linewidth]{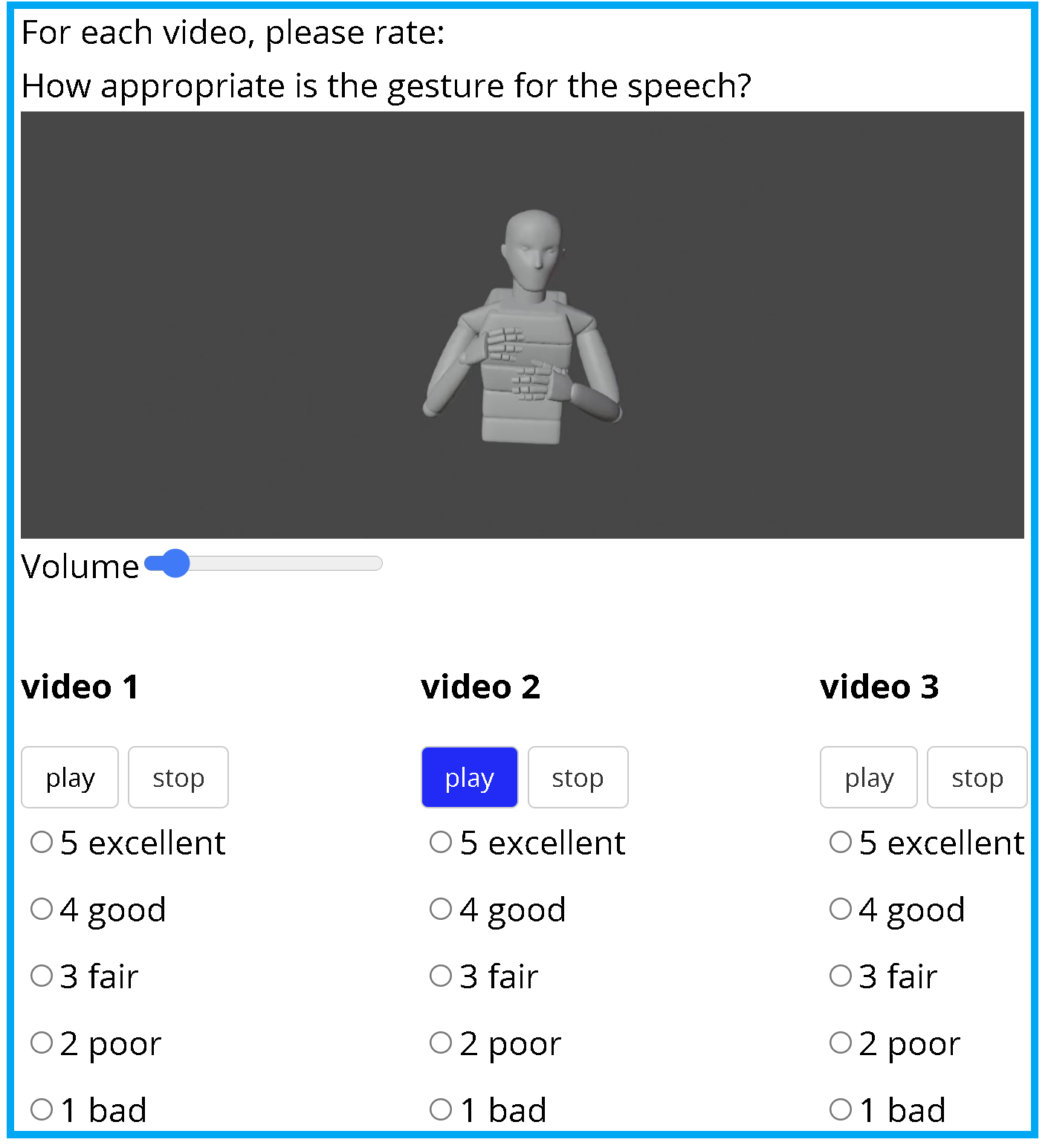}
   \vspace{-2em}
  \caption{Rendered video and interface for evaluation.}
  \Description{.}
  \label{fig:mushra_page}
\end{figure}

\subsection{Perceptual evaluation method}
We evaluate the proposed models in three separate tests: speech-and-gesture, gesture-only, and speech-only.
Assessing how models perform in each modality is necessary as users may be biased towards one modality when rating speech-and-gesture generation,
as demonstrated in a recent study focused on gesture-generation evaluation \cite{kucherenko2021large}.
It showed that mismatched ground-truth gesture, i.e., ground-truth gesture taken from a different speech segment, receives better scores than model-generated gestures on this dataset, likely because the motion itself, albeit unrelated to the speech, is more natural than model-generated motion.
Evaluating both modalities separately also allows us to understand how each modality contributes to the overall success of a model, and whether or not a model is biased towards either modality, for example generating better speech at the expense of worse gesture.

We render the generated gesture on a skinned avatar \cite{kucherenko2021large} with time-aligned speech at 20 fps.
We choose 20 fps because it is the lowest frame rate among the models.
All models are trained at frame rates that they perform best at (for both speech and gesture).
A screenshot of the resulting rendered video is shown in Figure \ref{fig:mushra_page}.

We use a MUSHRA-like \citep{itu2015method} (MUltiple Stimuli with Hidden Reference and Anchor) interface commonly used for subjective evaluation of speech-synthesis \cite{ribeiro2015perceptual}, but here adapted for video interfaces, since such setups have been found to work well for evaluating head motion and hand gestures \cite{braude2016head,jonell2021hemvip,kucherenko2021large}.
On a single test page, participants are presented with videos of generated gesture-speech from all evaluated models on the same input text sentence.
They can play in any order for any number of times, and rate each video based on a test question, on a standard MOS scale \cite{itu1996telephone}.
The order of the test pages and the order of videos on each page is independently randomized for each user.
This setup is used for both video-based tests (speech-and-gesture and gesture-only).
The speech test uses a similar MUSHRA-like interface.
A screenshot of the video-evaluation interface is shown in Figure \ref{fig:mushra_page}.
We recruit three separate groups of native English speakers for the three tests on the Prolific crowdsourcing platform.

\section{Perceptual evaluation results}
\label{sec:results}
\subsection{Speech-and-gesture evaluation}
\begin{figure}[b]
  \centering
  \includegraphics[width=\linewidth]{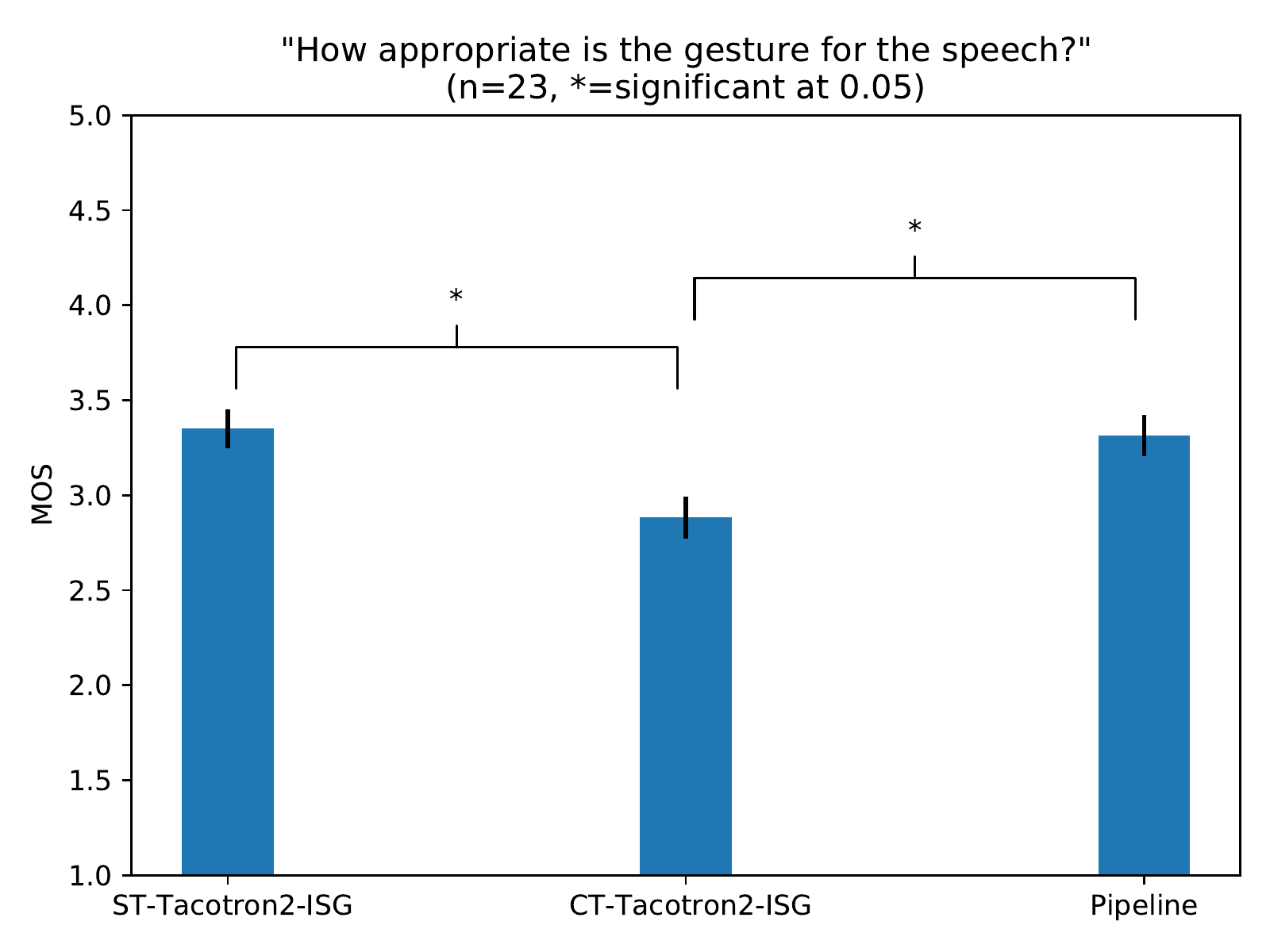}
   \vspace{-3em}
  \caption{Speech-and-gesture evaluation result.}
  \Description{.}
  \label{fig:co-speech-gesture_MOS}
\end{figure}

We asked 23 users to rate ``How appropriate is the gesture for the speech?'' for each model-generated speech-gesture pair.
This question is taken from \cite{kucherenko2021large} and is intended to assess the coherence between gesture and speech in general, including synchrony and meaningfulness.
GlowTTS-ISG is not evaluated in this test, nor in the speech-only test, since the generated speech quality does not approach the intelligibility standards necessary for meaningful perceptual evaluation.
The results are shown in Figure \ref{fig:co-speech-gesture_MOS}.
ST-Tacotron2-ISG obtains the highest MOS at 3.35 while the pipeline system scores 3.31, however the difference is not statistically significant.\footnote{Unless otherwise noted, statistical significance is tested using pairwise $t$-tests at $p$=0.05.}
We note that ST-Tacotron2-ISG achieves comparable performance to the pipeline despite being much more parameter efficient in its gesture generation.
CT-Tacotron2-ISG which updates both the speech and gesture sub-networks simultaneously scores lower than other models (statistically significant). 

\subsection{Gesture-only evaluation}
In this evaluation we ask participants to rate ``How human-like is the gesture?'' for each model-generated gesture.
This question is also taken from \cite{kucherenko2021large}.
It assesses how closely the generated gesture motion resembles ground-truth motion.
The videos used in this test are the same as the ones in the speech-and-gesture evaluation with GlowTTS-ISG videos added, but with audio turned off in order to remove the effect of the speech when assessing motion.

\begin{figure}[!tb]
  \centering
  \includegraphics[width=\linewidth]{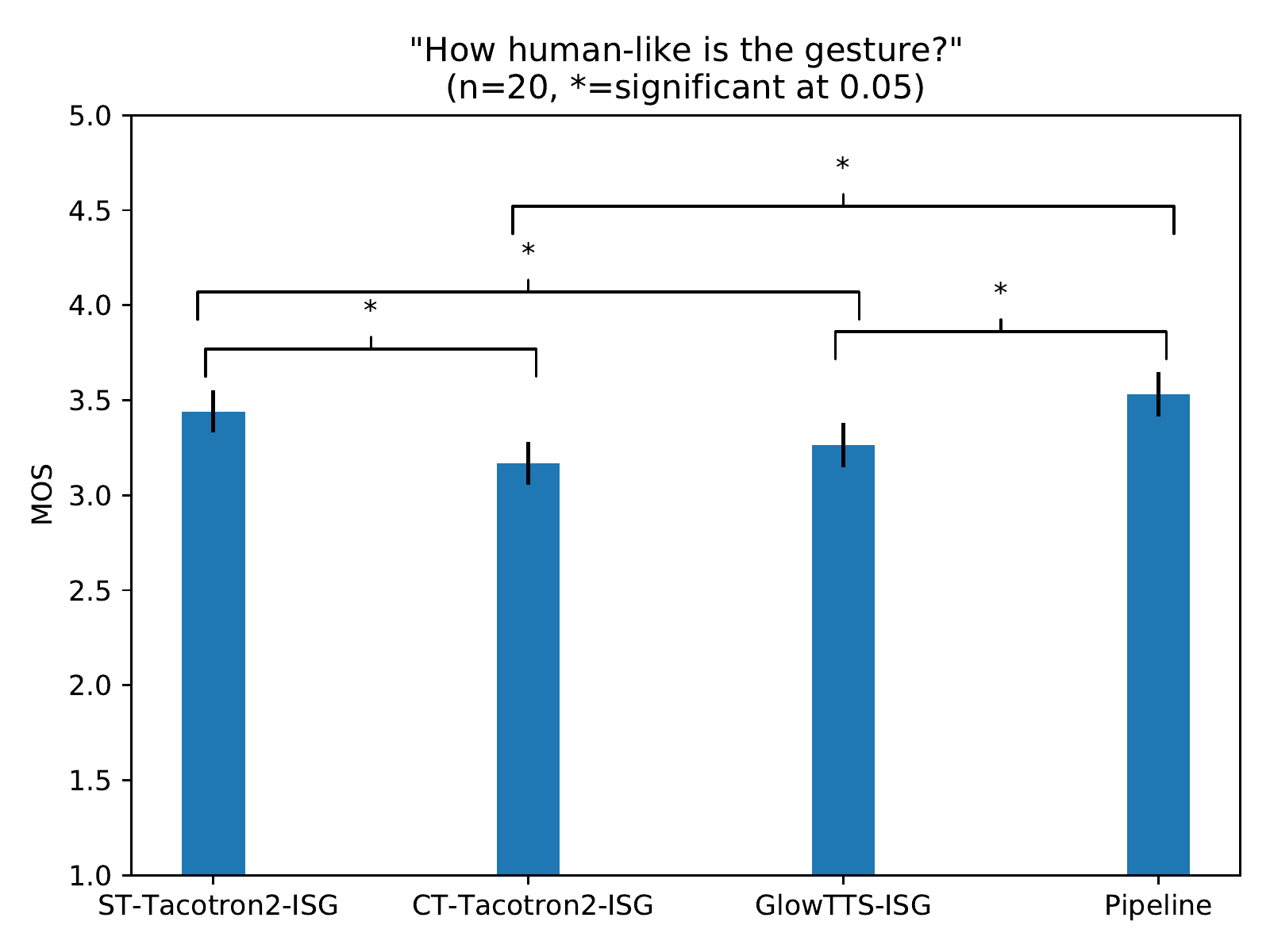}
 \vspace{-2.5em}
  \caption{Gesture-only evaluation result.}
  \Description{.}
  \label{fig:gesture_MOS}
\end{figure}

Figure \ref{fig:gesture_MOS} shows the result of the gesture-only evaluation.
The pipeline system (StyleGestures \cite{alexanderson2020style}) scores highest at 3.53 while ST-Tacotron2-ISG scores second best at 3.44, however the difference between the two is not statistically significant.
GlowTTS-ISG and CT-Tacotron2-ISG score lowest. Their difference is not statistically significant.
StyleGestures is significantly better than these two models while ST-Tacotron2-ISG is only significantly better than CT-Tacotron2-ISG.
We find StyleGestures to be very dynamic and having more detailed motion, such as subtle head bobbing when talking fast.
This is consistent with it scoring highest in this evaluation.
On the other hand, ST-Tacotron2-ISG is comparable to StyleGestures while having 3.5 times less parameters.

\begin{figure}[!t]
  \centering
  \begin{minipage}[c]{0.15\linewidth}
  \vspace{-3.5em}
      \textbf{ST-Tacotron2-ISG}
  \end{minipage}%
  \begin{minipage}[b]{0.6\linewidth}
      \centering
      \includegraphics[width=.5\linewidth]{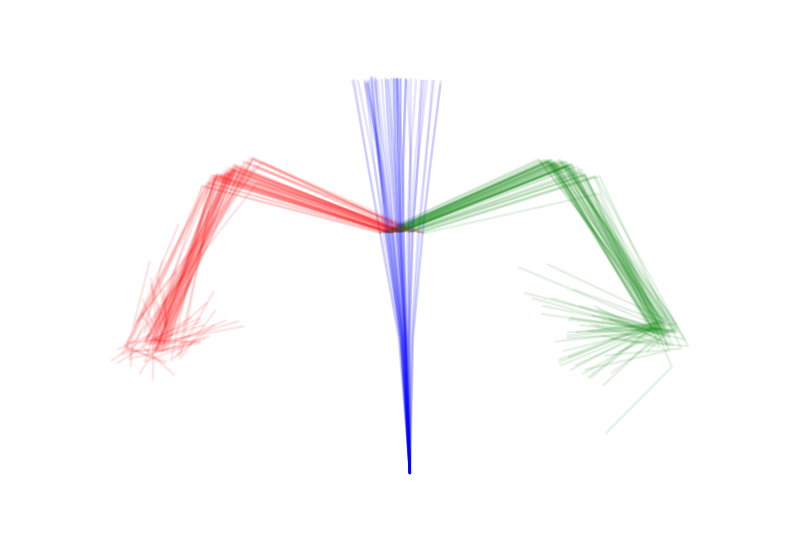}
  \end{minipage}%
  \begin{minipage}[b]{0.6\linewidth}
      \centering
      \hspace{-10em}
      \includegraphics[width=.5\linewidth]{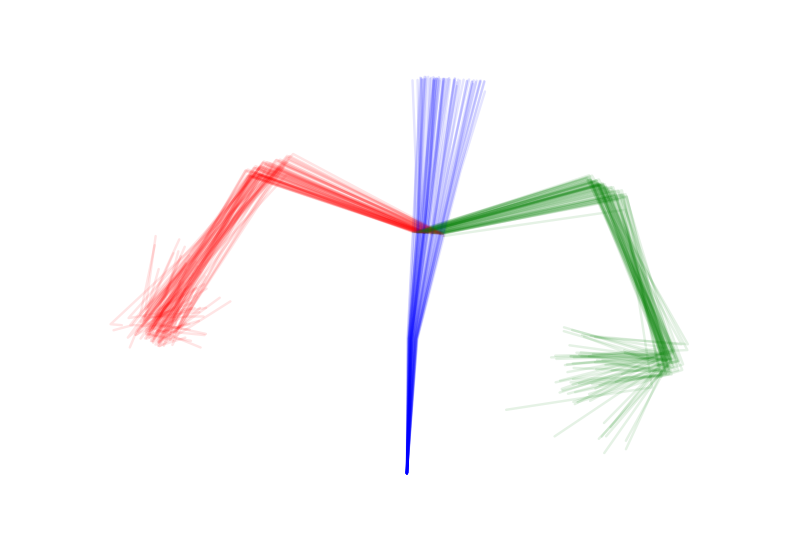}
  \end{minipage}%
  \\
  \rule{\linewidth}{0.4pt}
  \\
  \begin{minipage}[c]{0.15\linewidth}
   \vspace{-3.5em}
     \textbf{CT-Tacotron2-ISG}
  \end{minipage}%
  \begin{minipage}[b]{0.6\linewidth}
      \centering
      \includegraphics[width=.5\linewidth]{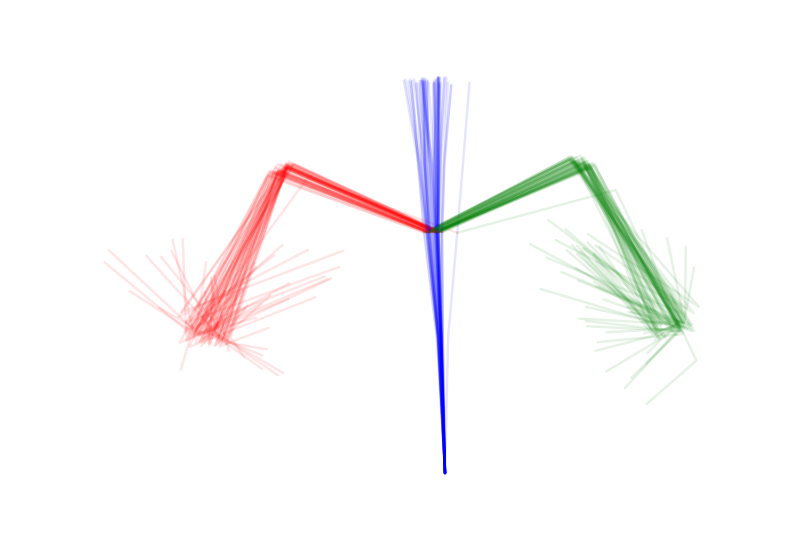}
  \end{minipage}%
  \begin{minipage}[b]{0.6\linewidth}
      \centering
      \hspace{-10em}
      \includegraphics[width=.5\linewidth]{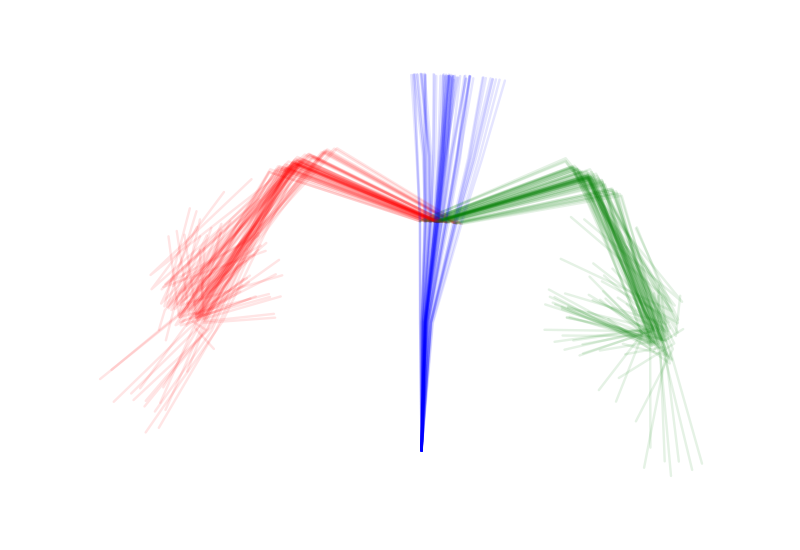}
  \end{minipage}%
  \\
  \rule{\linewidth}{0.4pt}
  \\
  \begin{minipage}[c]{0.15\linewidth}
   \vspace{-3.5em}
      \textbf{GlowTTS-ISG}
  \end{minipage}%
  \begin{minipage}[b]{0.6\linewidth}
      \centering
      \includegraphics[width=.5\linewidth]{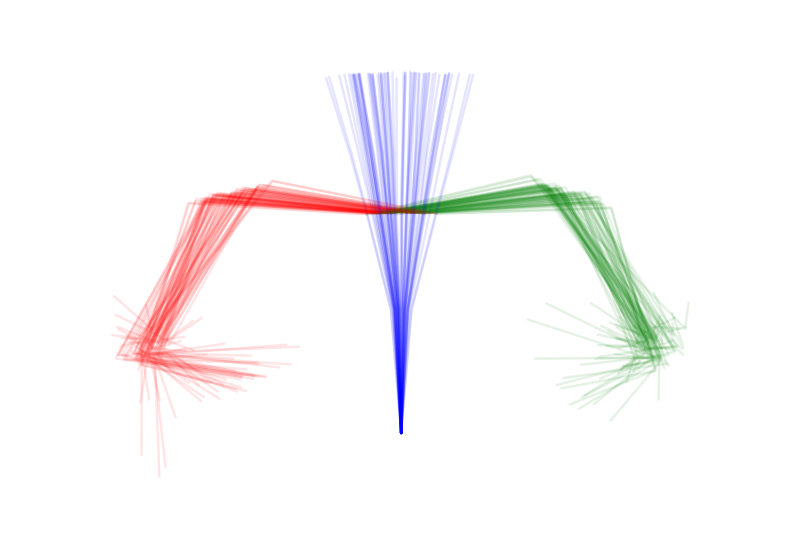}
  \end{minipage}%
  \begin{minipage}[b]{0.6\linewidth}
      \centering
      \hspace{-10em}
      \includegraphics[width=.5\linewidth]{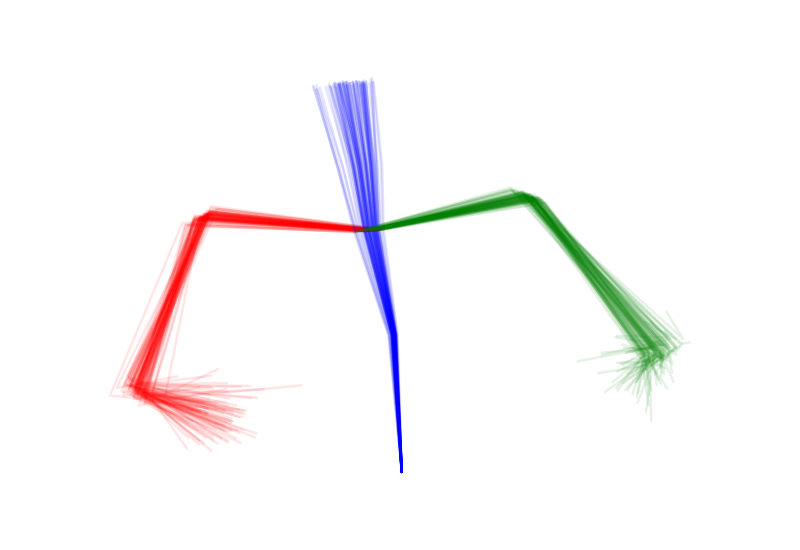}
  \end{minipage}%
  \\
  \rule{\linewidth}{0.4pt}
  \\
  \begin{minipage}[c]{0.15\linewidth}
   \vspace{-3.5em}
      \textbf{Pipeline\\(Style\-Gestures)}
  \end{minipage}%
  \begin{minipage}[b]{0.6\linewidth}
      \centering
      \includegraphics[width=.5\linewidth]{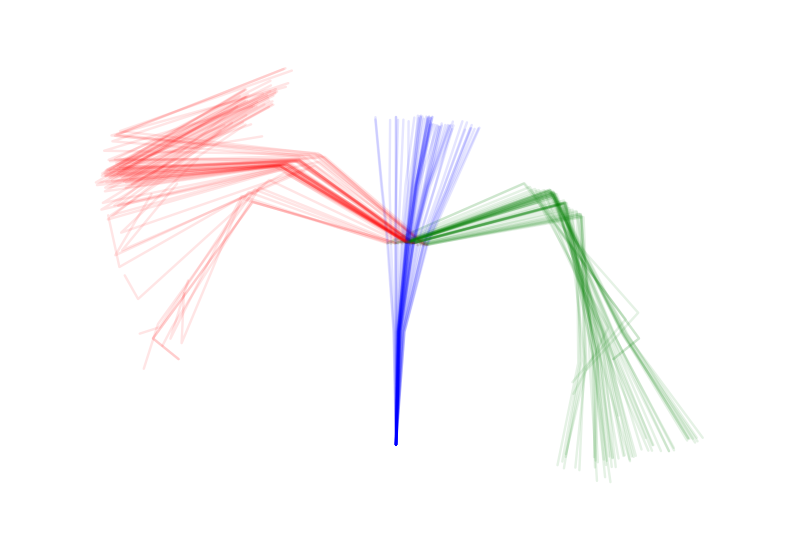}
  \end{minipage}%
  \begin{minipage}[b]{0.6\linewidth}
      \centering
        \hspace{-10em}
      \includegraphics[width=.5\linewidth]{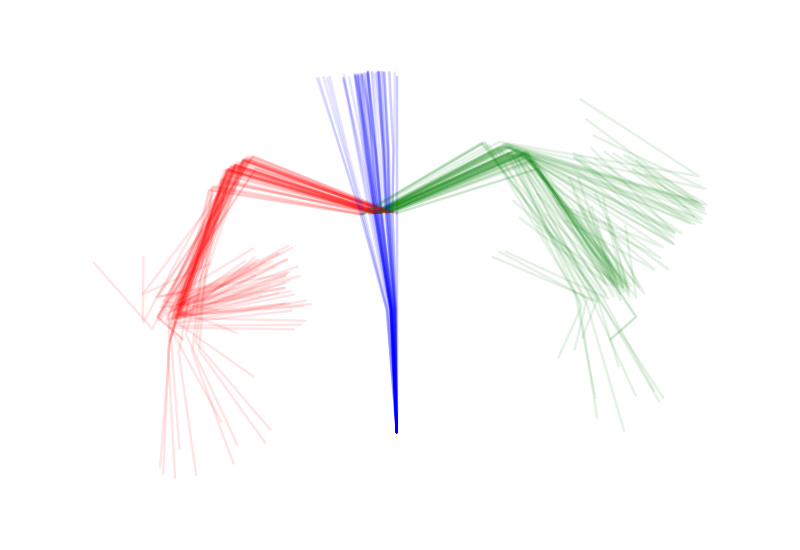}
  \end{minipage}%
  \\
   \hspace{+3em}
  \begin{minipage}{0.02\linewidth}
      \textbf{\mbox{Test~input~0}}
  \end{minipage}%
  \hspace{+10.5em}
  \begin{minipage}{0.02\linewidth}
      \textbf{\mbox{Test~input~1}}
  \end{minipage}%
  \caption{Visualizations of generated gesture space. Colors distinguish right arm (red), left arm (green), and torso (blue).}
  \label{fig:gestures}
\end{figure}

Figure \ref{fig:gestures} visualizes two evaluated gesture sequences generated by each of the 4 models for the same two inputs. 
All models learn to generate plausible gesture shapes. StyleGestures (row 4 in the figure) has the greatest range and variation, which is consistent with it receiving the highest rating in the gesture-only evaluation.

 \vspace{-0.5em}

\subsection{Speech-only evaluation}
The TTS community currently largely relies on questions similar to ``How natural does the speech sound?'' to evaluate synthesized speech.
However, some of the models we evaluate have speech errors such as skipped words, an aspect that we also want listeners to evaluate.
We thus combine naturalness and intelligibility evaluation and ask listeners to ``Please rate the synthesized speech audios based on a combination of: a) whether or not you can clearly hear each word, and b) how natural they sound.''
The input text is shown to users in this test, which is not the case for the other two tests. 

The models compared in this test are different than the other two tests.
We test three versions of Tacotron 2 to understand how ISG training affects the generated speech.
The three versions are, (a) full ISG from scratch (no speech-only pre-training), (b) speech-only training, and (c) ISG fine-tuning after speech-only training. All three are trained for the same number of iterations.
The speech-only model is the same as the speech sub-network of ST-Tacotron2-ISG and the pipeline system, which is just Tacotron 2 by itself.
GlowTTS-ISG is not evaluated in this test due to its low speech quality.

\begin{figure}[t]
  \centering
  \includegraphics[width=\linewidth]{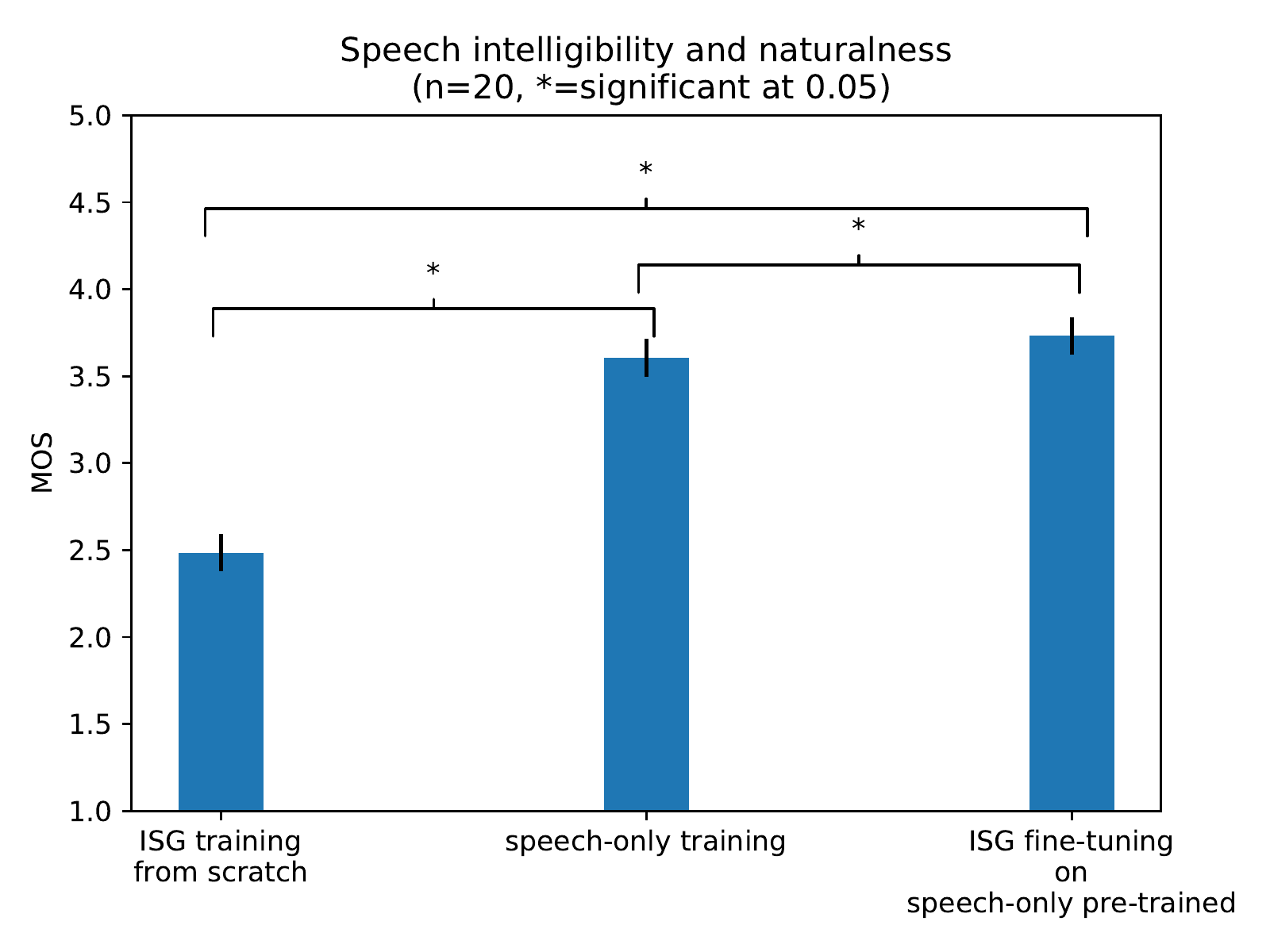}
   \vspace{-2.5em}
  \caption{Speech evaluation result.}
  \Description{.}
  \label{fig:speech_MOS}
\end{figure}

The results are shown in Figure \ref{fig:speech_MOS}.
ISG fine-tuning on top of speech-only training obtains the highest score at 3.62, better than both other training setups.
This could be due to increased training time, since the ISG system fine-tuned on top of the speech-only model, but even if that is the case, it still establishes that ISG can sound equally good as speech-only systems.
Full ISG training from scratch obtains the lowest score at 2.49, significantly worse than the second best speech-only training which obtains 3.49.
We think this is because the database size is smaller than what Tacotron 2 typically needs to create high-quality TTS, showing the benefits of a transfer-learning approach that leverages unimodal data for ISG.

 \vspace{-0.5em}

\section{Discussion}
ST-Tacotron2-ISG scores higher than CT-Tacotron2-ISG in both speech-and-gesture and gesture-only evaluations.
The biggest difference between the two models is that CT-Tacotron2-ISG trains both speech and gesture sub-networks together while ST-Tacotron2-ISG freezes the speech sub-network during gesture training.
This suggests that full ISG training, where both sub-networks are optimized simultaneously, may result in worse gesture as the overall model focuses on improving speech.
The speech loss could be given a reduced weight to balance the two modalities. Moreover, weighting in general is a potential way to improve overall synthesis quality by changing which modality the model focuses on during a certain training phase.
The speech-only evaluation reveals that transfer learning from larger, unimodal databases may be used to boost ISG quality and convergence time in both modalities. This is particularly appealing for ISG approaches, since speech data is much more widely available than aligned speech and 3D motion-capture gesture material.

One aspect of the proposed Tacotron2-ISG models we are particularly interested in is how the Tacotron 2 attention-layer informs gesture generation.
To probe this, we trained a similar model to Tacotron2-ISG in which the gesture sub-network takes the generated mel-spectrogram from Tacotron 2 as input, instead of the output from Tacotron 2 attention-layer.
We found that the model generated less articulated gestures than Tacotron2-ISG.\footnote{See \href{https://swatsw.github.io/isg_icmi21/}{https://swatsw.github.io/isg\_icmi21/} or supplemental material for video examples.}
This suggests that the Tacotron 2 attention layer provides features that facilitate gesture generation and that are not exposed by a pipeline approach.
However, other speech synthesis-frameworks may process information differently, and may be even more suited for an ISG approach.

We also observed that the proposed Tacotron2-ISG model is able to reproduce common speech-gesture patterns such as a subtle shrug and symmetrical hand gestures when saying ``I don't know''.
While the model itself does not have semantic input, the dataset contains several occurrences of ``I don't know'' in different contexts, and it is possible that the gesture sub-network has learned to associate the attention-layer representation of that phrase with the shrugging gestures in the database.

Furthermore, the poor synthetic speech generated by GlowTTS-ISG does not imply that a probabilistic generative model cannot achieve good ISG.
The issues may at least partially be explained by database size, since normalizing flows endeavor to learn the dual-modal data distribution, whereas minimizing the MSE only requires being able to predict its mean.
However, the strong performance of unimodal flow-based systems such as StyleGestures show that these models still have plenty of potential for ISG.

\section{Limitations}
As presented here, integrated speech and gesture generation relies on a database of text, audio, and 3D motion in parallel. Such databases are smaller and less common than databases containing only a subset of these modalities.
Furthermore, we only investigated two TTS models; other TTS models may also be suitable for the task.
Lastly, evaluation of speech and gesture remains challenging for the research community in general, and innovations in evaluation approaches could reveal additional nuances of the tested models.

\section{Conclusions}
We introduce integrated speech and gesture generation (ISG), where both modalities are generated jointly in a single architecture, as a new research problem that brings together TTS and gesture generation.
We propose several models for ISG and evaluate them in a set of carefully designed user studies, on each modality separately and on both modalities combined.
Taken together, the results from all three studies demonstrate that one of the proposed ISG models (ST-Tacotron2-ISG) performs comparably to the current state-of-the-art pipeline system, while being faster and having much fewer parameters. Our findings, and the challenges we identified along the way, suggest that ISG is a promising and largely unexplored topic which deserves attention from synthesis researchers across communities.

\begin{acks}
This research was supported by Swedish Research Council projects 2019-05003 (Connected) and 2018-05409 (StyleBot), by Digital Futures (AAIS), the Riksbankens Jubileumsfond project P20-0298 (CAPTivating), and by the Wallenberg AI, Autonomous Systems and Software Program (WASP) funded by the Knut and Alice Wallenberg Foundation.
\end{acks}

\bibliographystyle{ACM-Reference-Format}
\bibliography{ibg-bib}

\end{document}